\documentclass[a4paper,twoside,10pt]{letter}
\usepackage{graphicx,saj,multicol}

\def\papertype{}

\def\udc{}
\begin{document}

\baselineskip=3.1truemm
\columnsep=.5truecm


\markboth{\eightrm Possible thermal emission at radio frequencies from SNR HB 3}
{\eightrm D. Oni{\' c} and D. Uro{\v s}evi{\'c}}

{\ }

\publ

\type

{\ }


\title{An analysis of the possible thermal emission at radio frequencies from an evolved supernova remnant HB 3 (G132.7$+$1.3): revisited}


\authors{D. Oni{\' c}$^{1}$ and D. Uro{\v s}evi{\'c}$^{1}$}

\vskip3mm


\address{$^1$ Department of Astronomy, Faculty of Mathematics,
University of Belgrade, Studentski trg 16, 11000 Belgrade, Serbia}

\Email{donic}{matf.bg.ac}{yu}


\dates{}{}


\summary{It has recently been reported that some of the flux
density values for an evolved supernova remnant (SNR) HB 3
(G132.7$+$1.3) are not accurate. In this work we revised an
analysis of the possible thermal emission at radio frequencies
from the SNR HB 3 using the recently published, corrected, flux
density values. A model including a sum of non-thermal (purely
synchrotron) and thermal (bremsstrahlung) component is applied for
fitting integrated radio spectrum of the SNR. The contribution of
thermal component in total volume emissivity at $1\ \mathrm{GHz}$
was estimated to be $\approx37 \%$. The ambient density was also
estimated to be $n\approx 9\ \mathrm{cm}^{-3}$ for the
$\mathrm{T}=10^{4}\ \mathrm{K}$. Again, we obtained the relatively
significant presence of thermal emission at radio frequencies from
the SNR so we could support interaction between SNR HB 3 and
adjacent molecular cloud associated with the \mbox{H\,{\sc ii}}
region W3. Our model estimates for thermal component contribution
to total volume emissivity at $1\ \mathrm{GHz}$ and ambient
density are similar to those obtained earlier ($\approx40\ \%$,
$\approx10\ \mathrm{cm^{-3}}$). It is clear that the corrected
flux density values do not change the basic conclusions.}



\begin{multicols}{2}
{


\section{1. INTRODUCTION}

The presence of thermal emission at radio frequencies may be a
useful tool for identifying interactions between supernova
remnants (SNRs) and molecular clouds, and also for estimating the
ambient density near SNRs using radio continuum data (Uro{\v s}evi{\' c} \& Pannuti 2005, Uro{\v s}evi{\' c}, Pannuti \& Leahy 2007).
In this work we argue for the presence of a thermal bremsstrahlung component in the radio emission from SNR HB 3 in addition to the synchrotron component.
It is possible that SNRs can be sources with significant amount of
thermal radiation and as Uro{\v s}evi{\' c} \& Pannuti (2005) stated, there are two basic
criteria for the production of a significant amount of radio
emission through the thermal bremsstrahlung process from an SNR:
the SNR evolves in an environment denser than the average and its
temperature must be lower than the average (but always greater
then the recombination temperature). Two cases have been
considered: thermal emission at radio frequencies from a
relatively young SNR evolving in a dense molecular cloud
environment ($n\approx100-1000\ \mathrm{cm^{-3}}$) and extremely
evolved SNR (approximately $10^{5}-10^{6}$ years old) expanding in
a dense warm medium ($n\approx1-10\ \mathrm{cm^{-3}}$). For a detailed discussion about the
issue see Uro{\v s}evi{\' c} \& Pannuti (2005) and Uro{\v s}evi{\' c}, Pannuti \& Leahy (2007).

Uro{\v s}evi{\' c}, Pannuti \& Leahy (2007) have analyzed the broadband ($22-3900\ \mathrm{MHz}$)
radio spectrum of the Galactic SNR HB 3 and have discussed the
possible thermal radio emission from the SNR. Earlier published
observations have revealed that a curvature is present in the
radio spectrum of SNR HB 3, indicating that a single synchrotron
component appears insufficient to adequately fit the radio
spectrum. Uro{\v s}evi{\' c}, Pannuti \& Leahy (2007) have suggested that more natural explanation
for the apparent spectral index variations found by Tian \& Leahy (2005) is
synchrotron emission, which dominates at lower frequencies, and
bremsstrahlung emission, which dominates at higher frequencies.
They have found $\approx40\ \%$ for the thermal component
contribution to total volume emissivity at $1\ \mathrm{GHz}$ and
have also estimated the ambient density, implied by the presence
of thermal component, to be $\approx10\ \mathrm{cm^{-3}}$.
Recently, Green (2007) has reviewed the radio spectrum of SNR HB 3,
noting the difficulties in deriving accurate flux densities for
the remnant, particularly at higher frequencies, due to thermal
emission from the nearby bright \mbox{H\,{\sc ii}} region W3 (IC
1795) and its surroundings. He pointed out that some of the
earlier published flux density values used by Uro{\v s}evi{\' c}, Pannuti \& Leahy (2007) in
their analysis are not accurate and that the spectrum of the SNR
is well represented by a simple power low spectrum as well as the
contamination with thermal emission from adjacent regions is the
cause for the reported spectral flattening of the spectrum. In
this work we present the results of our analysis using the
corrected, recently reported by Green (2007), flux densities for SNR
HB 3.

\section{2. The model}

A spectrum of SNRs in radio domain is usually represented by an
ordinary power law. If the frequency is in GHz, the flux density
can be represented by the following expression:
\begin{equation}
S_{\nu}=S_{1\ \!\!\mathrm{GHz}}\cdot\nu^{-\alpha},
\end{equation}
where: $S_{1\ \!\!\mathrm{GHz}}$ is the flux density at 1 GHz, and
$\alpha$ is the radio spectral index. In order to distinguish the
contribution of thermal and non-thermal component within the total
radiation, SNR radio integrated spectrum was fitted by a simple
sum of these two components. If the frequencies are still in GHz,
the relation for the flux density can be written as follows:
\begin{equation}
S_{\nu}=S_{1\ \!\!\mathrm{GHz}}^{\mathrm{NT}}\ (\nu^{-\alpha}+\frac{S_{1\ \!\!\mathrm{GHz}}^{\mathrm{T}}}{S_{1\ \!\!\mathrm{GHz}}^{\mathrm{NT}}}\ \nu^{-0.1}),
\end{equation}
where: $S_{1\ \!\!\mathrm{GHz}}^{\mathrm{T}}$ and $S_{1\
\!\!\mathrm{GHz}}^{\mathrm{NT}}$ are flux densities corresponding
to thermal and non-thermal component, respectively. The spectral
index is considered to be constant in the SNR shell. It is also
considered that the thermal radiation is optically thin and has
the spectral index equal 0.1 at any point. As the radio frequency
increases, the amount of synchrotron radiation from an SNR
decreases and the amount of thermal bremsstrahlung emission
becomes more significant. In our model it is also considered that
the synchrotron radiation, optically thin at any point, is not
absorbed or scattered by thermal gas.

The volume emissivity of thermal bremsstrahlung radiation for an
ionized gas cloud is proportional to the square of the electron
(or ion) volume density $n$, i.e.:
\begin{equation}
 \varepsilon_{\nu}=7\times10^{-38}n^{2}T^{-\frac{1}{2}},
\end{equation}
where $n$ is in $\mathrm{cm}^{-3}$ and thermodynamic temperature
$T$ in K. Having determined total $\varepsilon_{\nu}$ and thermal
component contribution to total volume emissivity, the density of
the of the interstellar medium (ISM) can be estimated using
equation 3.

This model is valid only in the approximation of constant density
and temperature. The model itself also presumes a simple sum of
non-thermal and thermal component, while the attention is not paid
to the fact that the dependence of flux density could be another,
more complicated, function of thermal and non-thermal components.
It is also important to note that this model, in general, does not
distinguish between thermal and non-thermal emission with the same
spectral index (i.e.\@ the case of lower synchrotron spectral
index).

Despite these drawbacks, our model represents a useful tool for
estimating the contribution of thermal bremsstrahlung component to
the total volume emissivity and ambient density using radio
continuum data.
}
\end{multicols}

\vfill\eject

\begin{multicols}{2}
{

\section{3. SNR HB 3 (G132.7+1.3)}

From the Green (2006) paper, the value of the radio spectral index
is 0.4, $S_{1\ \!\!\mathrm{GHz}}=45\ \mathrm{Jy}$, size is around
80 arcmin and the SNR is S (shell) type. On the other hand, a
combination of radio shell morphology with a center-filled thermal
X-ray morphology has led to the classification of SNR HB 3 as a
mixed-morphology SNR (Uro{\v s}evi{\' c}, Pannuti \& Leahy 2007
and references therein). It is the one of the largest SNRs
currently known (Fesen et al. 1995, Tian \& Leahy 2005). The
distance from the SNR is about $2.2\ \mathrm{kpc}$ (Tian \& Leahy
2005, Shi et al. 2008). SNR HB 3 size, based on a distance of $2\
\mathrm{kpc}$ is $60\times80\ \mathrm{pc}$ (Tian \& Leahy 2005 and
references therein). Lazendi\'{c} \& Slane (2006) stated that the
SNR is $90\times120\ \mathrm{arcmin}$ in diameter. Kovalenko,
Pynzar \& Udal'tsov (1994) have reported: $\alpha=(0.51\pm0.12)$,
Fesen et al. (1995): $\alpha=(0.64\pm0.01)$ (also pointed by
Lazendi\'{c} \& Slane 2006) and Landecker et al. (1987):
$\alpha=(0.60\pm0.04)$. Green (2007) has found
$\alpha=(0.56\pm0.03)$. Tian \& Leahy (2005) pointed spectral
index variations with most values between 0.3 and 0.7.

Shi et al. (2008) have extracted 4800 MHz total intensity and
polarization data of HB 3 from the Sino-German 6 cm polarization
survey of the Galactic plane made with the Urumqi 25 m telescope,
but they could not give a total flux density at 4800 MHz of the
whole SNR because of a low resolution. They have found a radio
spectral index of HB 3 of $\alpha = -(0.61\pm0.06)$ using only
three flux densities, at 1408 MHz, 2695 MHz and 4800 MHz, and have
concluded that there is no spectral flattening at high
frequencies. Shi et al. (2008) have also pointed out that a reliable
observations of SNR HB 3 at frequencies above 3000 MHz are crucial
to confirm a spectral flattening.

A radio pulsar PSR J0215+6218 has been discovered within (in
projection) the SNR HB 3 boundary, but it appears to be much older
than the remnant and therefore not associated with the SNR
(Lazendi\'{c} \& Slane 2006, Lorimer, Lyne \& Camilo 1998).

Fesen et al. (1995) pointed that SNR HB 3 is relatively optically faint
SNR. Diffuse and filamentary optical emission has been detected
from the SNR, with the strongest emission along the western SNR
shell (Lazendi\'{c} \& Slane 2006). Reich, Zhang \& F\"{u}rst (2003) pointed that the SNR radio
shell is located at the western edge of the \mbox{H\,{\sc ii}}
region complex W3-W4-W5. Optical emission from the SNR was found
to be well-correlated with the radio emission, with a multiple
shock structure found in the western SNR shell and lack of
emission in the southeast region (Lazendi\'{c} \& Slane 2006 and references
therein).

Most of the mixed-morphology SNRs are interacting with molecular
or \mbox{H\,{\sc i}} clouds, as indicated in some cases by
infrared line emission or OH masers (Rho \& Petre 1998). OH (1720 MHz)
masers, which are recognized as a diagnostic for a molecular cloud
interaction with a SNR, have been detected towards the W3/HB 3
complex (Lazendi\'{c} \& Slane 2006, Koralesky et al. 1998).

Uro{\v s}evi{\' c}, Pannuti \& Leahy (2007) pointed that the X-ray emission is seen to lie
entirely within the radio shell of HB 3. To obtain an independent
estimate of the ambient density of the ISM surrounding HB 3, they
have performed spectral fitting on the extracted {\textit ASCA}
GIS spectra and have calculated electron densities to be
$n_{e}\approx0.4f^{-1/2}\ \mathrm{cm^{-3}}$ for the central region
and $n_{e}\approx0.1f^{-1/2}\ \mathrm{cm^{-3}}$ for the northern
and southern regions (see Table 2 in Uro{\v s}evi{\' c}, Pannuti \& Leahy 2007), where $f$
represents the volume filling factor.

\section{Analysis and results}

Green (2007) pointed out that the radio observations of the SNR HB 3
are complicated by confusing thermal emission from an adjacent
\mbox{H\,{\sc ii}} region W3. He has shown that some of the
earlier published flux density values (used by Uro{\v s}evi{\' c}, Pannuti \& Leahy 2007 in
their analysis) are not accurate. Green (2007) also pointed the huge
problem of deriving the accurate flux density values and has
listed the corrected values for SNR HB 3. Particulary, he reported
corrected values for 408 and 1420 MHz points and also corrected
uncertainty for 865 MHz point, all from Tian \& Leahy (2005) and references
therein. He also derived an integrated flux density for the
Effelsberg 2695 MHz survey data. Green (2007) excluded 3650 and 3900
MHz points (from Tian \& Leahy 2005 and references therein) due to
their possible contamination with thermal emission associated with
W3. He also rescaled 22, 38 and 178 MHz points, used by Uro{\v s}evi{\' c}, Pannuti \& Leahy (2007) in their analysis, to be on the scale of Baars et al. (1977).
We have revised an analysis of the possible thermal emission
contribution in total volume emissivity at radio frequencies from
an evolved SNR HB 3 using data points for integrated radio flux
density from Green (2007) for a range of 22 MHz to 2.695 GHz (see
Table 1 in Green 2007).

The parameters of our model fit (equation 2) are as follows:
$\alpha=(0.70\pm0.05)$, $S_{1\
\!\!\mathrm{GHz}}^{\mathrm{NT}}=(30.45\pm5.22)$ Jy, $S_{1\
\!\!\mathrm{GHz}}^{\mathrm{T}}/S_{1\
\!\!\mathrm{GHz}}^{\mathrm{NT}}=(0.59\pm0.27)$,
$\chi_{\mathrm{red}}^{2}=0.19$, 9 degrees of freedom (dof) as it
can be seen in Table 1. $\chi_{\mathrm{red}}^{2}$ represents
reduced $\chi^{2}$ ($\chi^{2}/\mathrm{dof}$). The parameters of
purely non-thermal model fit (equation 1) are as follows:
$\alpha=(0.56\pm0.02)$, $S_{1\
\!\!\mathrm{GHz}}^{\mathrm{NT}}=(49.18\pm2.06)$ Jy,
$\chi_{\mathrm{red}}^{2}=0.31$, 10 dof as it can be seen in Table
2.

}
\end{multicols}

\vfill\eject

\begin{multicols}{2}
{

{\bf Table 1.} The fit parameters for our model for SNR HB3.
\vskip2mm
\centerline{\begin{tabular}{c c c c}
\hline
$\alpha$ & $S_{1\ \!\!\mathrm{GHz}}^{\mathrm{NT}}\ \mathrm{(Jy)}$ & $\frac{S_{1\ \!\!\mathrm{GHz}}^{\mathrm{T}}}{S_{1\ \!\!\mathrm{GHz}}^{\mathrm{NT}}}$ & $\chi_{\mathrm{red}}^{2}\ \mathrm{(dof)}$\\
\hline
$0.70\pm0.05$ & $30.45\pm5.22$ & $0.59\pm0.27$ & 0.19 (9)\\
\hline
\end{tabular}}

\vskip.5cm

{\bf Table 2.} The fit parameters for purely non-thermal model for SNR HB3.
\vskip2mm
\centerline{\begin{tabular}{c c c}
\hline
$\alpha$ & $S_{1\ \!\!\mathrm{GHz}}^{\mathrm{NT}}\ \mathrm{(Jy)}$ & $\chi_{\mathrm{red}}^{2}\ \mathrm{(dof)}$\\
\hline
$0.56\pm0.02$ & $49.18\pm2.06$ & 0.31 (10)\\
\hline
\end{tabular}}

\vskip.5cm

It can be noted that the radio spectral index value is higher than
the value from Green (2006) both for the purely non-thermal
($\alpha=0.56\pm0.02$) and our model fit ($\alpha=0.70\pm0.05$)
calculations. The results from the purely non-thermal model fit
responds to the values placed in Green (2007). Our model fit radio
spectral index estimate is closer to the value placed in
Lazendi\'{c} \& Slane (2006) and Fesen et al. (1995) than in Green
(2006, 2007). Figure 1 shows in a solid line fit by non-thermal plus thermal
model, while the dotted line shows the fit by purely non-thermal
model through the given sample.

\centerline{\includegraphics[height=8cm]{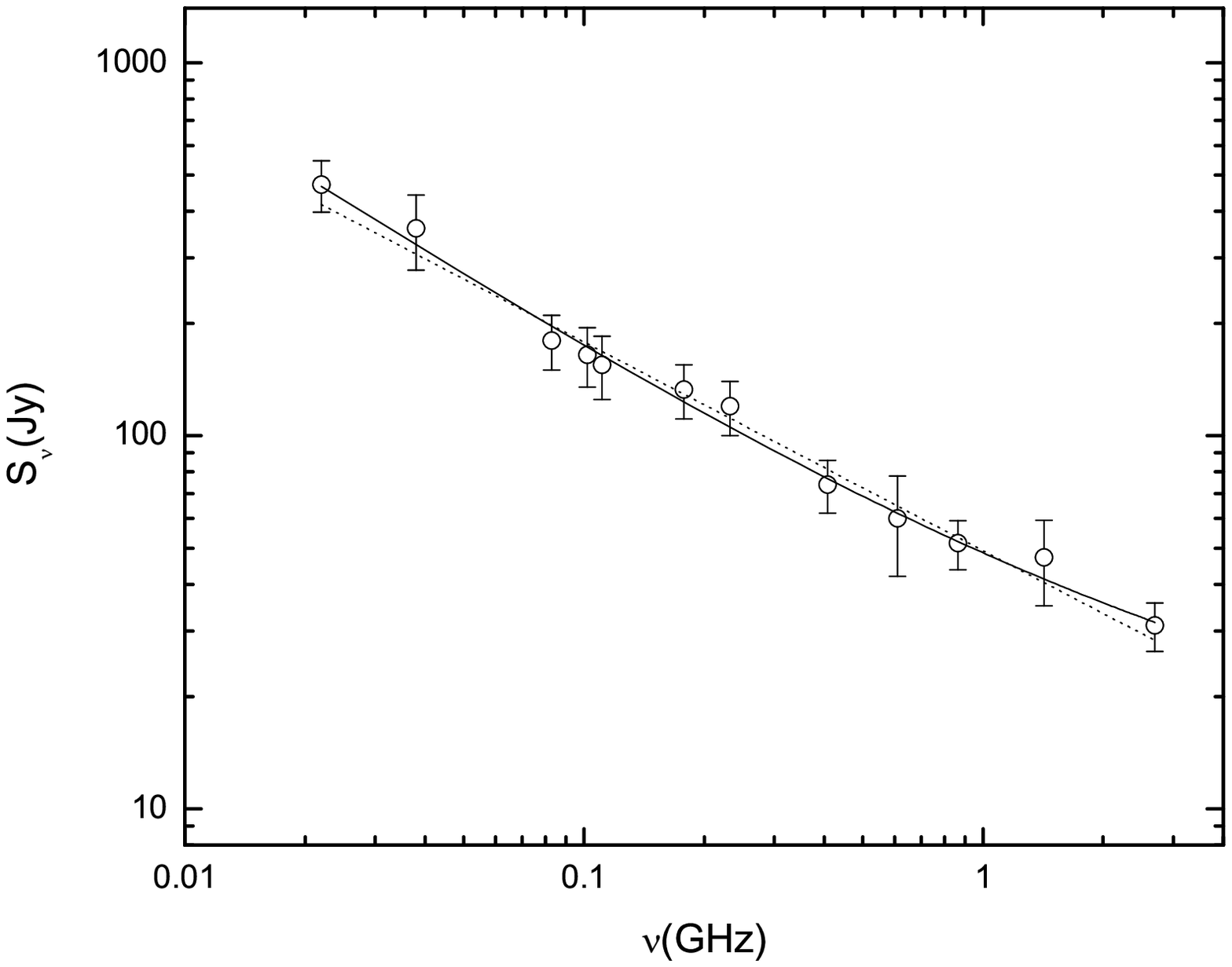}}

\figurecaption{1.}{The integrated spectrum of HB 3. The full line represents the fit by non-thermal plus thermal model, while the dotted line represents fitting by purely non-thermal model.}

\vskip.5cm

If a mean value of $37\ \%$, from our analysis, for the thermal
component contribution to total volume emissivity at $1\
\mathrm{GHz}$ is assumed, we get $n\approx9\ \mathrm{cm}^{-3}$ for
the assumed post-shock temperature value of $10^{4}\ \mathrm{K}$.
We have assumed $d=2\ \mathrm{kpc}$ and $D=70\ \mathrm{pc}$ for
consistency with Uro{\v s}evi{\' c}, Pannuti \& Leahy (2007). It
is clearly seen that for the corrected integrated flux density
values we also get the relatively significant presence of thermal
emission at radio frequencies from the SNR. Our model estimates
for thermal component contribution to total volume emissivity at
$1\ \mathrm{GHz}$ ($\approx37\ \%$) and ambient density
($\approx9\ \mathrm{cm^{-3}}$) are similar to those obtained
earlier ($\approx40\ \%$, $\approx10\ \mathrm{cm^{-3}}$) by Uro{\v
s}evi{\' c}, Pannuti \& Leahy (2007). The fact that essentially
the same thermal component again minimizes the $\chi^{2}$,
indicates that its presence cannot be ruled out by the corrections
to the flux densities.

If we assume the value for the compression parameter to be 4 we
can roughly estimate pre-shock ISM number density as
$n_{0}\approx2.25\ \mathrm{cm}^{-3}$ for $T=10^{4}\ \mathrm{K}$.

It is clearly visible that our ambient density estimates could
support the possibility that the SNR is indeed expanding in a
dense ISM. Based on our analysis we can support the possibility
that SNR HB 3 is indeed interacting with the molecular cloud
material. The presence of the thermal bremsstrahlung component in
the radio spectrum of SNR HB 3 suggests that this SNR is in fact
interacting with adjacent molecular cloud associated with the
\mbox{H\,{\sc ii}} region W3.

It should be pointed out that the further measurements at the
highest radio frequencies ($>3\ \mathrm{GHz}$) are required for a
detailed discussion of the issue.

\section{Conclusions}

In this work we revised an analysis of the possible thermal
emission at radio frequencies from an evolved SNR HB 3. Some of
the earlier published flux density values for SNR HB 3 are shown
to be non-accurate. Here we present the results of our analysis
using the recently published, corrected, flux densities so the
main conclusions are:

\renewcommand{\theenumi}{(\arabic{enumi})}

\item {1.} The contribution of thermal component in total volume emissivity was estimated to be $\approx37 \%$ and the ambient density was also estimated to be
$n\approx 9\ \mathrm{cm}^{-3}$ for the $\mathrm{T}=10^{4}\ \mathrm{K}$.

\item {2.} Our model estimates for thermal component
contribution to total volume emissivity at $1\ \mathrm{GHz}$ and
ambient density are similar to those obtained earlier ($\approx40\
\%$, $\approx10\ \mathrm{cm^{-3}}$). It is clear that the
corrected flux density values do not change the basic conclusions.

\item {3.} The presence of the thermal bremsstrahlung component in the radio spectrum of SNR HB 3
suggests that this SNR is in fact interacting with adjacent
molecular cloud associated with the \mbox{H\,{\sc ii}} region W3.
The presence of thermal emission at radio frequencies may be a
useful tool for identifying interactions between SNRs and
molecular clouds and also for estimating the ambient density near
SNRs using the radio continuum data.

\item {4.} The lack of data at higher radio frequencies unable us to give a firmer conclusion about the issue.

\acknowledgements{We acknowledge Dave Green for useful comments. This paper is a part of the projects No. 146012 and No. 146016 supported by the Ministry of Science of Serbia.}

\vskip.5cm

\references

Baars, J. W. M., Genzel, R., Pauliny-Toth, I. I. K., Witzel, A., 1977, \journal{A\&A}, \vol{61}, 99

Fesen, R. A., Downes, R. A., Wallace, D., Normandeau, M., 1995, \journal{AJ}, \vol{110}, 2876

Green, D. A., 2007, \journal{BASI}, \vol{35}, 77

Green, D. A., 2006, A catalogue of Galactic Supernova Remnants (2006 April version). Astrophysics Group, Cavendish Laboratory, Cambridge, http://www.mrao.cam.ac.uk/surveys/snrs/

Koralesky, B., Frail, D. A., Goss, W. M., Claussen, M. J., Green, A. J., 1998, \journal{AJ}, \vol{116}, 1323

Kovalenko, A. V., Pynzar, A. V., Udal'tsov, V. A., 1994, \journal{ARep}, \vol{38}, 95

Landecker, T. L., Vaneldik, J. F., Dewdney, P. E., Routledge, D., 1987, \journal{AJ}, \vol{94}, 111

Lazendi\'{c}, J. S, Slane, P. O., 2006, \journal{ApJ}, \vol{647}, 350

Lorimer, D. R., Lyne, A. G., Camilo, F., 1998, \journal{A\&A}, \vol{331}, 1002

Reich, W., Zhang, X., F\"{u}rst, E., 2003, \journal{A\&A}, \vol{408}, 961

Rho, J., Petre, R., 1998, \journal{ApJ}, \vol{503}, L167

Routledge, D., Dewdney, P. E., Landecker, T. L., Vaneldik, J. F., 1991, \journal{A\&A}, \vol{274}, 529

Shi, W. B., Han, J. L., Gao, X. Y., Sun, X. H., Xiao, L., Reich, P. and Reich, W., 2008, \journal{A\&A}, arXiv:0806.1647

Tian, W., Leahy, D., 2005, \journal{A\&A}, \vol{436}, 187

Uro{\v s}evi{\' c} D, Pannuti T. G., 2005, \journal{Astropart. Phys.}, \vol{23}, 577

Uro{\v s}evi{\' c} D, Pannuti T. G., Leahy D., 2007, \journal{ApJ}, \vol{655}, L41

\endreferences

}
\end{multicols}

\vfill\eject

{\ }



\naslov{ANALIZA MOGU{\CC}E TERMALNE EMISIJE NA RADIO FREKVENCIJAMA EVOLUIRANOG OSTATKA HB3}


\autori{D. Oni\cc$^{1}$ i D. Uroxevi{\cc}$^{1}$}

\vskip.5cm


\adresa{$^1$Katedra za astronomiju, Matematiqki fakultet, Univerzitet u
Beogradu\break Student\-ski trg 16, 11000 Beograd, Srbija}

\vskip.7cm


\centerline{UDK \udc}


\centerline{\rit Prethodno saopxtenje}

\vskip.7cm

\begin{multicols}{2}
{


{\rrm {Nedavno je objavljeno da su neke od vrednosti gustina fluksa evoluiranog ostatka supernove (OSN)}} HB 3 (G132.7$+$1.3) {\rrm{neta{\ch}ne. U ovom radu smo ponovili analizu mogu{\cc}e termalne radio-emisije OSN}} HB 3 {\rrm{ koriste{\cc}i nedavno publikovane, korigovane vrednosti, gustina fluksa. Model koji pretpostavlja sumu netermalne i termalne komponente je primenjen za fitovanje radio-spektra ostatka. Prisustvo termalne komponente u ukupnoj zapreminskoj emisivnosti na $1\ \mathrm{GHz}$ je procenjeno na $\approx37\ \%$. Gustina okolne sredine je takodje procenjna na $n\approx 9\ \mathrm{cm}^{-3}$ za $\mathrm{T}=10^{4}\ \mathrm{K}$. Ponovo je odredjeno zna{\ch}ajno prisustvo termalne komponente u ukupnoj zapreminskoj emisivnosti tako da mo{\zz}emo da podupremo hipotezu o interakciji izmedju ostatka i molekularnog oblaka. Procene prisustva termalne komponente u ukupnoj zapreminskoj emisivnosti na $1\ \mathrm{GHz}$ i gustine okolne sredine su sli{\ch}ne sa ranije odredjenim ($\approx40\ \%$, $\approx10\ \mathrm{cm^{-3}}$). Jasno je vidljivo da vrednosti korigovanih gustina fluksa ne me{\nj}aju osnovne zaklju{\ch}ke.}}
}
\end{multicols}

\end{document}